\documentclass[%
paper=a4, %
twocolumn, %
english, %
DIV=16,
headings=normal,
headinclude=false,%
footinclude=false,%
]{scrartcl}%

\usepackage[english]{babel}
\usepackage{ftnright}
\usepackage{amsmath}
\usepackage{graphicx}
\usepackage{amsmath}
\usepackage{bm}
\usepackage{siunitx}
\usepackage[colorinlistoftodos]{todonotes}
\usepackage{textcomp}

\addtokomafont{title}{\raggedright}
\addtokomafont{author}{\raggedright\normalsize}
\addtokomafont{date}{\raggedleft\normalsize}

\addtokomafont{captionlabel}{\sffamily\bfseries}%
\addtokomafont{caption}{\sffamily\footnotesize}

\setcapindent{0em}

\usepackage[backend=bibtex,style=nature, isbn=false]{biblatex}
\usepackage[colorlinks=true, allcolors=blue]{hyperref}

\newbibmacro{string+doi}[1]{%
	\iffieldundef{doi}{#1}{\href{http://dx.doi.org/\thefield{doi}}{#1}}}
\DeclareFieldFormat{title}{\usebibmacro{string+doi}{\mkbibemph{#1}}}
\DeclareFieldFormat[article]{title}{\usebibmacro{string+doi}{\mkbibquote{#1}}}

\newcommand{\um}{\,\textmu m}
\newcommand{\El}{\ensuremath{E_0}}  
\newcommand{\Ekin}{\ensuremath{E_{\text{kin}}}}  
\newcommand{\Elexp}{\ensuremath{E_0^\prime}}  

\newcommand{\Eoff}{\ensuremath{E_{\text{off}}}}  
\newcommand{\dOff}{\ensuremath{d_{\text{off}}}}  
\newcommand{\ds}{\ensuremath{d_{\text{s}}}}  

\newcommand{\Vl}{\ensuremath{V_0}}  
\newcommand{\VMM}{\ensuremath{V_{\text{MMT}}}}  
\newcommand{\Vs}{\ensuremath{V_{\text{s}}}}  

\newcommand{\WF}{\ensuremath{\Phi}}  
\newcommand{\WFx}{\ensuremath{\Phi(x)}}  
\newcommand{\dWF}{\ensuremath{\Delta\Phi}}  
\newcommand{\dWFapp}{\ensuremath{\Delta\Phi_{\text{app}}}}  
\newcommand{\dx}{\ensuremath{\Delta x}}  

\newcommand{\Ev}{\ensuremath{E_{\text{vac}}}}  
\newcommand{\EF}{\ensuremath{E_{\text{F}}}}  

\newcommand{\ei}{\ensuremath{I}}  
\newcommand{\IMM}{\ensuremath{I_{\text{MM}}}}  
\newcommand{\ILEEM}{\ensuremath{I_{\text{LEEM}}}}  

\newcommand{\Iexp}{\ensuremath{I_{\text{exp}}(x_i, \Elexp)}}  
\newcommand{\Isim}{\ensuremath{I_{\text{sim}}(x_i, \El, \dOff)}}  
\newcommand{\ssr}{\ensuremath{\text{ssr}(\Elexp, \dOff)}}  
\newcommand{\SSR}{\ensuremath{\text{SSR}(\Eoff, \dOff)}}

\newcommand{\tio}{TiO\textsubscript{2}}
\newcommand{\sro}{SrO}

\newcommand{\fig}[1]{Fig.\ \ref{fig:#1}}

\begin{document}

\title{Quantifying work function differences using low-energy electron microscopy: the case of mixed-terminated strontium titanate}

\author{
	Johannes Jobst\textsuperscript{{1}*},
	Laurens M.\ Boers\textsuperscript{{1}},
	Chunhai Yin\textsuperscript{{1}},
	Jan Aarts\textsuperscript{{1}}, \\ 
	Rudolf M.\ Tromp\textsuperscript{{2,1}},
	Sense Jan van der Molen\textsuperscript{{1}}
}

\date{Submitted on 07 December 2018}

\maketitle

\let\thefootnote\relax\footnote{%
	\textsuperscript{1}\,Leiden Institute of Physics, Leiden University, Niels Bohrweg 2, P.O. Box 9504, NL-2300 RA Leiden, The Netherlands.\\
	\textsuperscript{2}\,IBM T.J. Watson Research Center, 1101 Kitchawan Road, P.O. Box 218, Yorktown Heights, New York 10598, USA.\\
	* e-mail \href{mailto:jobst@physics.leidenuniv.nl}{jobst@physics.leidenuniv.nl}\\
}


\textbf{
	For many applications it is important to measure the local work function of a surface with high lateral resolution. Low-energy electron	microscopy is regularly employed to this end since it is, in principle, very well suited as it combines high resolution imaging with high sensitivity to local electrostatic potentials. For surfaces with areas of different work function, however, lateral electrostatic fields inevitably associated with work function discontinuities deflect the low-energy electrons and thereby cause artifacts near these discontinuities.
	We use ray-tracing simulations to show that these artifacts extend over hundreds of nanometers and cause an overestimation of the true work	function difference near the discontinuity by a factor of 1.6 if the standard image analysis methods are used. We demonstrate on a mixed-terminated strontium titanate surface that	comparing LEEM data with detailed ray-tracing simulations leads to much a more robust estimate of the work function difference.
}



\section{Introduction}\label{sec:introduction}	
\begin{figure}[t]
	\includegraphics[width=\columnwidth]{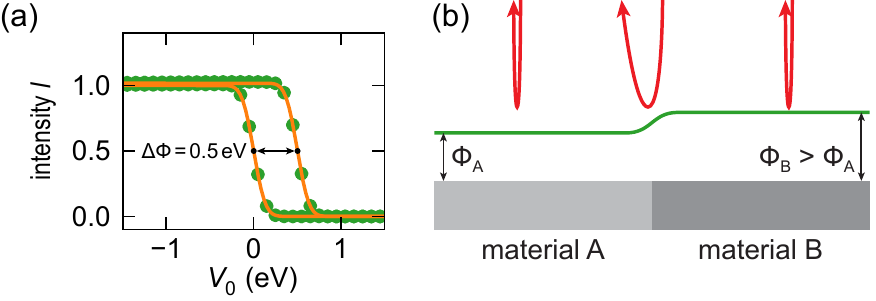}
	\caption{Sketch of the canonical way to determine \dWF\ from LEEM IV-curves and of the problem with WF-induced, in-plane fields.
		(a) The start voltage \Vl\ at which the mirror-mode transition occurs in IV-curves recorded at two different positions is shifted according to \dWF.
		(b) This is strictly valid only when electrons reflect from uniform surfaces (left and right trajectories). At the boundary between two materials with different WF, the potential landscape (green line) is deformed causing a deflection of the electron trajectory (center). This intrinsic effect strongly affects LEEM images and the extracted, apparent WF.}
	\label{fig:sketch}
\end{figure}

The work function (WF) \WF\ of a material is the energy needed to remove an electron from the surface into the vacuum, i.e., the difference between vacuum energy and Fermi level $\WF = \Ev - \EF$. It is an important fundamental property defining, for example, the photoemission threshold \cite{Fowler1931-photoemission} as well as the energy landscape when multiple materials are brought into contact. It is therefore of great technological relevance for photocathodes, thermionic emission  and band alignment in semiconductor devices such as high-k transistors or solar cells \cite{Sze2006}. 
Many materials can exhibit diverse surface reconstructions or terminations with different \WF\ depending on their treatment and often exhibit domains of different WF on the surface \cite{Minohara2010-termination, Aballe2015-instability, Back2017-WF, Cauduro2017-WF}. To understand those materials and to utilize them to their full potential, it is thus necessary to quantify local WF differences \dWF\ with high lateral resolution. 
Low-energy electron microscopy (LEEM) is, in principle, ideally suited to map out \dWF\ because the electron landing energy \El\ can be adjusted precisely. This is achieved by decelerating the electrons that leave the objective lens with a kinetic energy of 15\,keV to energies close to zero by lifting the sample to a potential of $-15\,\text{kV} + \Vl$.
By increasing the start voltage \Vl, we can, thus, slowly go from mirror mode ($\El < 0$), where all electrons are reflected before they reach the surface, to LEEM mode ($\El > 0$) where they interact with the material. This mirror-mode transition (MMT) is accompanied by a steep drop of the reflected electron intensity \ei\ in the so-called IV-curve (intensity vs.\ start voltage \Vl) as sketched in \fig{sketch}(a). 
The inflection point of the MMT is a precise measure of the condition $\El = 0$ and thus, the vacuum energy \Ev\ since it identifies the energy at which electrons from the vacuum (LEEM probing electrons) start to interact with the surface. 
As \EF\ is constant throughout the sample, WF differences cause a shift of the energetic position of the MMT. 
Measuring this shift of the MMT for all IV-curves in an area is widely used to extract local WF differences due to the good lateral resolution of LEEM \cite{Schramm-thesis} and its sensitivity to small \dWF\ \cite{Kennedy2011-Laplacian, Kautz-LEEP, Jobst-LEEP}. 

This sensitivity, however, poses major challenges in the interpretation of the data that are typically overlooked in literature. In particular, at the boundary between two materials with different WF, the equipotential lines above the surface are curved, as sketched in \fig{sketch}(b). The resulting lateral electrical fields deflect trajectories of the low-energy electrons on their way towards the sample as well as after reflection. 
These deflections create imaging artifacts that can easily lead to misinterpretation of data.
This effect is particularly problematic for the extraction of \dWF\ from mirror mode shifts since the electron energy is close to zero around the MMT, thus causing the largest deflection-induced artifacts. 

In this paper, we show that these effects are unavoidable around WF discontinuities and quantify the magnitude and lateral size of the resulting artifacts in the determined \dWF\ by ray-tracing simulations. We lay out a methodology to extract correct values of \dWF\ by comparing experimental data with simulation results. We demonstrate this framework on the mixed-terminated surface of strontium titanate (STO) but it is applicable to virtually any material and geometry.

\begin{figure}[t]
	\includegraphics[width=\columnwidth]{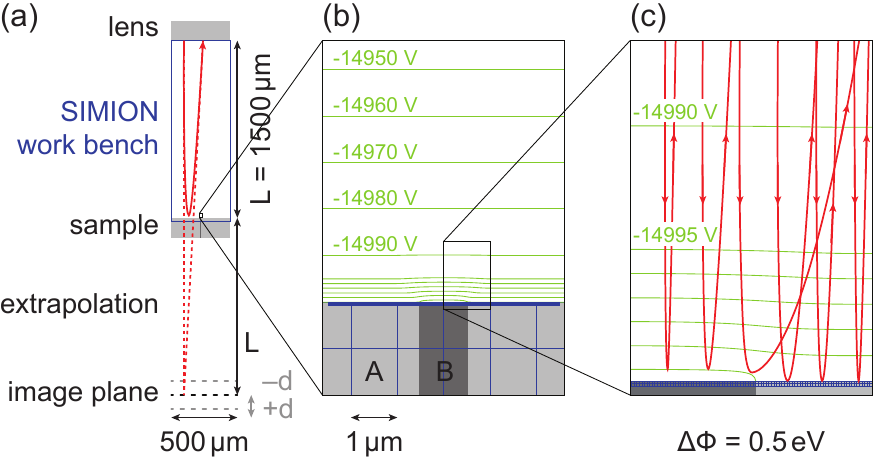}
	\caption{Electron trajectories are simulated in the cathode lens region using SIMION \cite{Dahl2000-SIMION}. 
		(a) The potentials between objective lens (grounded) and sample (at $-15\,\text{kV} + \Vl$) are calculated on a 1\um\ grid. In the vicinity of the sample, the resolution is improved by using a nested 10\,nm grid. The returning trajectories are extrapolated linearly to a virtual image plane at $2L + d$ to form an image with defocus $d$. 
		(b) Zoom into the surface region showing both grids (blue) and calculated equipotential lines (green) for $\Vl = 0$.
		(c) A further shows zoom how the \dWF\ between material A and B (light to dark gray, respectively) deforms the equipotential lines (green) and thus, deflects the simulated electron trajectories (red).}
	\label{fig:sim-geometry}
\end{figure}

\section{Setup of the simulations}\label{sec:simsetup}

To simulate LEEM images arising from a WF discontinuity, we perform ray-tracing simulations using the program SIMION 8.1.1.32 \cite{Dahl2000-SIMION}. We reproduce the experimental geometry of the cathode lens region and simulate the trajectories of a collimated, incoming beam of electrons and their deflection due to \dWF. 
We set up the simulation with the parameters of the cathode lens in the ESCHER setup \cite{ESCHER} (based on a Specs FE-LEEM P90 AC \cite{Tromp-AC1, Tromp-AC2}), i.e., a distance of $L = 1.5$\,mm between sample and objective lens and a sample potential of $\Vs = -15\,\text{kV} + \Vl$. 
To keep simulation times manageable while being able to resolve fine lateral details in the sample, we use two nested work benches in SIMION as shown in \fig{sim-geometry}. The outer one with a grid size of 1\um\ and 1500\um\ $\times$  500\um\ in size, and the inner one close to the sample surface with a finer grid size (10\,nm for the single WF step and 5\,nm for the more complex geometry, see below) and a size of 20\um\ $\times$ 10\um.
We simulate an electron beam of width $b$ (3\um\ and 4\um\ for the two geometries) by calculating the trajectories of $n=2001$ electrons starting at the objective lens with $\Ekin = 15$\,keV and equal spacing. 
We record their position and velocity when they return to the objective lens after reflection from the sample and then  reconstruct an image by projecting them back to the virtual image plane behind the sample at $2L$, as sketched in \fig{sim-geometry}(a). Images defocused by $d$ are simulated by projecting to $2L + d$ (underfocused and overfocused conditions correspond to positive and negative $d$, respectively). 
Note that this is only valid as we omit relativistic effects that rescale the relation between kinetic electron energy and speed in the simulated trajectories, thus causing slightly different angles at the lens. Moreover, we do not consider the shift of the focus plane with landing energy since this effect (200\,nm shift per eV) is negligibly small for the energies considered here.
The local image intensity is given by the ratio between the incoming electron intensity and the spacing $\Delta r(x)$ between adjacent rays in this plane as described in Ref.\ \cite{Kennedy2011-Laplacian}:
\begin{equation}
	I(x, \El, d) = \frac{b/n}{\Delta r(x, \El, d)}.
\end{equation}
We take the non-ideal lateral resolution in mirror mode and the energy spread of the electrons into account by smoothing those simulated images with a Gaussian of standard deviation $\sigma_x = 10$\,nm and one with $\sigma_E = 0.11$\,eV, respectively. 
The sample consist of two materials, A and B, that exhibit a WF difference of \dWF\, which is simulated by off-setting the potential of the respective areas by \dWF.
Identical to the experimental situation, we vary the total sample potential in the simulation to change the landing energy of the electrons. We reference $\El = 0$ with respect to material A, the large part of the sample shown in lighter gray in \fig{sim-geometry}).

In the following, we discuss simulation results of two sample geometries. First, we study a single stripe of material A embedded in a homogeneous area of material B. Second, we simulate a geometry that we observe experimentally in a mixed-terminated STO sample that consists of multiple, almost parallel stripes of alternating materials.

\section{Results \& Discussion}
\subsection{Quantifying work-function-induced artifacts}
\begin{figure}[!ht]
	\includegraphics[width=\columnwidth]{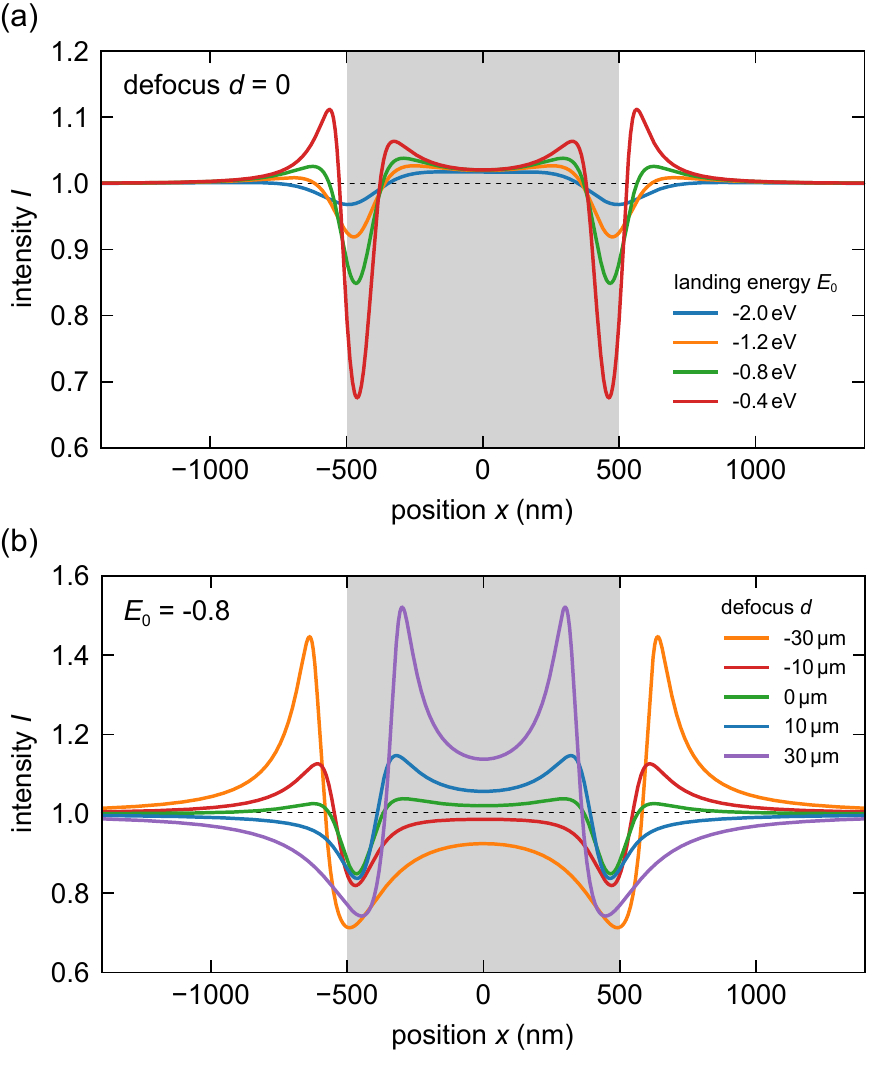}
	\caption{Simulated image intensity profiles $I(x)$ along a line across a stripe of different WF (gray area) of $\dWF = 0.5$\,eV.
		(a) Profiles for focused condition ($d = 0$) for various landing energies, referenced to the outer (white) areas. The reflectivity shows maxima and minima at the edge of the WF stripe. They correspond to bright and dark fringes in LEEM images close to the MMT. The features become more pronounced for slower electrons (closer to $\El = 0$).
		(b) Profiles for $\El = -0.8$\,eV for various focusing conditions. The shape and position of the features around the WF step strongly depends on defocus $d$, posing a great challenge to identify $d = 0$ in experiments.}
	\label{fig:sim-profile}
\end{figure}

In order to quantify imaging artifacts induced by WF changes, we start with the simplest geometry possible: a single stripe of material A with a WF of $\WF + \dWF$ is surrounded by material B with a WF of \WF. In particular, we simulate LEEM images around the MMT and determine the resulting errors in the extracted value of \dWF, 

Figure \ref{fig:sim-profile}(a) shows the simulated image contrast of a 1\um\ wide stripe (gray area) with $\dWF = 0.5$\,eV compared to its surrounding for different $\El < 0$ (counted with respect to the surrounding). 
For these MM images, the intensity is $I = 1$ far away from the work function discontinuity as it would be for a uniform sample. In the vicinity of the discontinuity, however, the image intensity exhibits pronounced maxima and minima, whose amplitudes increase with increasing \El\ (coming closer to the MMT).
They are a direct result of the caustics induced by the lateral fields and thus, affect slower electrons more strongly \cite{Kennedy-MM-distortions, Kennedy2011-Laplacian}.
 
In experimental LEEM images, these features correspond to dark and bright fringes that are regularly observed around work function discontinuities (e.g., Fig.\ 3(a) in Ref.\ \cite{Back2017-WF}). 
The fact that their contrast is energy-dependent, often makes the interpretation of images in mirror mode challenging \cite{Kennedy-MM-distortions}. 
Moreover, the appearance of these fringes strongly depends on the focusing conditions as shown in \fig{sim-profile}(b). Depending on the defocus $d$, some fringes can be enhanced or even completely suppressed. 
This poses a serious problem for experiments since the condition of perfect focus can thus not be identified straightforwardly \cite{Nataf2016-MEM-focus} as the image contains significant contrast even at $d = 0$.

\begin{figure}[!ht]
	\includegraphics[width=\columnwidth]{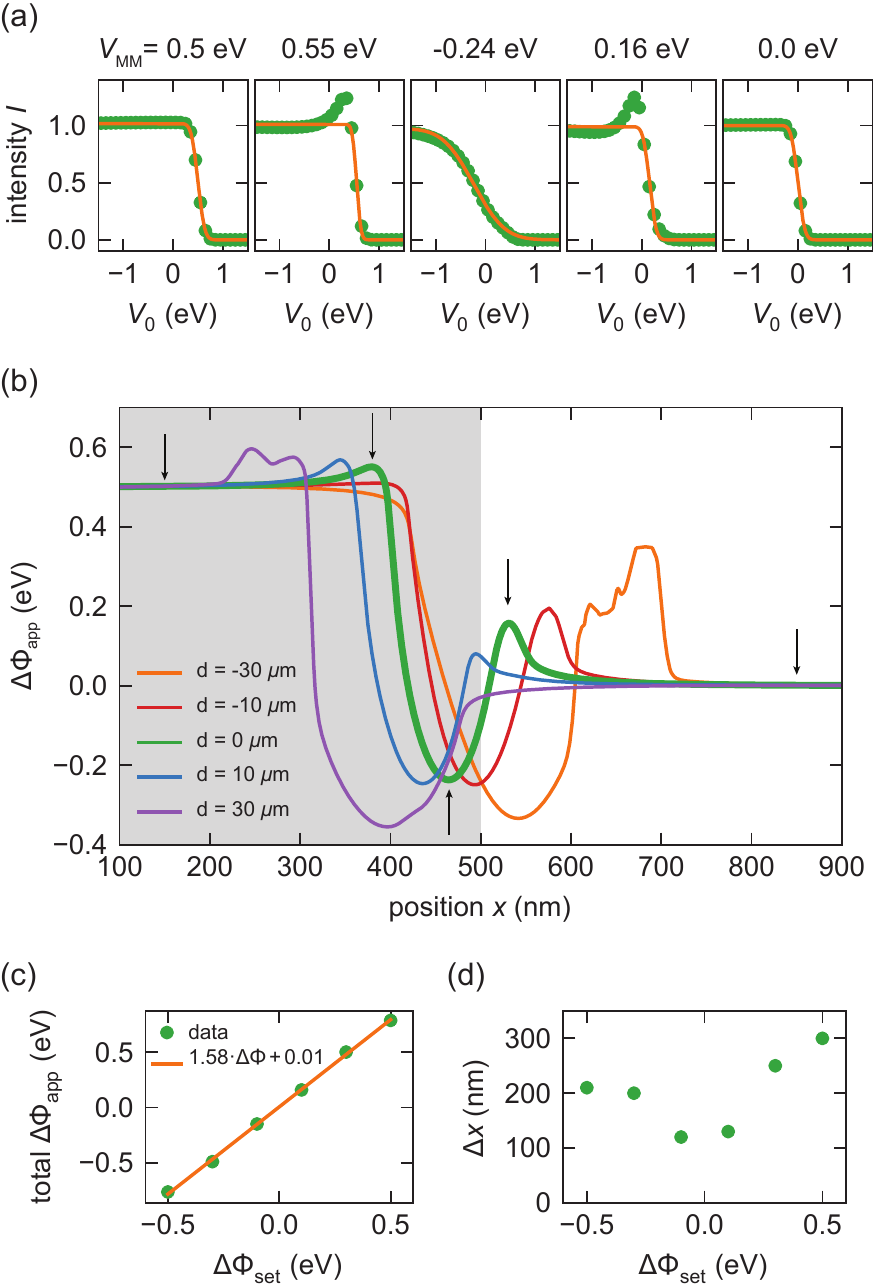}
	\caption{Work function discontinuities intrinsically affect the IV-curves and thus the extracted apparent WF difference \dWFapp.
		(a) IV-curves at different positions on the sample. Far away from the WF step, the curves can be well described by an error function and the resulting \dWFapp\ value is correct. In the vicinity of the discontinuity, the IV-curves show field-induced artifacts that affect the fitting result. 
		(b) The \dWFapp\ is  extracted as the position of the MMT from fitting Eq.\ (\ref{eq:error-function}) to the IV-curves at all positions along the line. Around the WF step, a pronounced artifact arises. Its exact shape depends on the focusing condition, but it is generally $\sim300$\,nm wide for $\dWF = 0.5$\,eV.
		(c) The extracted total \dWFapp\ (maximal minus minimal value of \dWFapp\ at $d = 0$) scales linearly with the true \dWF, i.e., it always overestimates the WF difference by a factor of $a \approx 1.6$.
		(d) The size of the WF artifact grows with increasing \dWF, but is already significant for very small WF differences.}
	\label{fig:sim-WF}
\end{figure}

To make matters worse, these WF-induced artifacts strongly affect the IV-curves and with this the extracted local work function as we will show in the following. 
We demonstrate this by deriving apparent WF differences \dWFapp\ from the simulations following the procedure that is generally used for experimental LEEM data, where local WF differences $\dWFapp(x)$ are identified as shifts of the MMT in IV-curves [cf.\ \fig{sketch}(a)] \cite{Cauduro2017-WF, Back2017-WF, Rault2012-thickness}. The exact voltage of the MMT \VMM\ is extracted by fitting an error function to the IV-curve:
\begin{equation}\label{eq:error-function}
I = \frac{\IMM - \ILEEM}{2} \cdot \text{erfc}\left(\frac{\Vl - \VMM}{\sqrt{2}\sigma}\right) + \ILEEM
\end{equation}
where
\begin{equation} 
\text{erfc}\left(x\right) = \frac{2}{\sqrt{\pi}} \int_{0}^{x} \text{e}^{-r^2} \text{d}r
\end{equation}
with \IMM\ and \ILEEM\ the intensities deep in mirror mode ($\El \ll 0$) and in LEEM mode ($\El \gg 0$), respectively and $\sigma$ the standard deviation describing the Gaussian spread of the electron energy.
For normalized data sets, Eq.\ (\ref{eq:error-function}) simplifies with $\IMM \approx 1$ and $\ILEEM \approx 0$ for many materials.

Since the simulations yield three-dimensional data sets $I(x, \El, d)$, they can either be visualized along the space coordinate as intensity cuts of an image $I(x)$ for different landing energies as in \fig{sim-profile}(a) or for different defocus values as in \fig{sim-profile}(b). Alternatively, they can be displayed along the energy coordinate as IV-curves $I(\El)$ for various points $x$ on the sample for a given $d$. 
Figure \ref{fig:sim-WF}(a) shows such simulated IV-curves for focused condition ($d = 0$) for five points around the WF discontinuity, which are indicated by arrows in \fig{sim-WF}(b). 
Far away from the WF discontinuity, the IV-curves show a clear drop from 1 to 0 and are well described by Eq.\ (\ref{eq:error-function}). Here, the extracted \VMM\ and $\sigma$ correspond exactly to the values that were used as input in the simulation. This indicates, that this method is reliable for uniform samples. 
Close to the point of WF change, on the other hand, the shape of the IV-curves strongly deviates from the expected behavior, showing either a peak before the MMT or a slow decrease of intensity instead of a sudden drop. 
The exact shape of the IV-curves is determined by the deflection of electron trajectories by local in-plane fields, which depends on the electron energy in a non-trivial way. 
Since the resulting IV-curves deviate from the simple step-like behavior, the description by an error function is no longer valid and the extracted \VMM\ by fitting Eq.\ (\ref{eq:error-function}) yields unphysical results that lie outside of the \dWF\ set in the simulation. 

Nevertheless, we use this method on simulated data to quantify the error it introduces in the extracted \dWFapp\ since this is the method canonically used for experimental data. Figure \ref{fig:sim-WF}(b) shows local WF differences extracted this way for different defocus values $d$. We find that \dWFapp\ clearly overshoots the true $\dWF = 0.5$\,eV around the WF step even at focused conditions (green, bold line). 
Moreover, the exact position, shape and amplitude of those artifacts strongly depends on focusing conditions. In particular, the pronounced minimum at the WF discontinuity is always present but its position and width changes with defocus, while both maxima in the region of high (gray) and low (white) work function can be removed by overfocusing or underfocusing, respectively. 
The comparison of Fig.\ 5(c) of Ref.\ \cite{Jobst-LEEP} with those simulations suggests that the data shown in Fig.\ 5 of Ref.\ \cite{Jobst-LEEP} was acquired at slightly overfocused conditions (negative $d$) as the additional maximum in the high \dWFapp\ region is absent. This shows that finding focusing condition, which is challenging in a LEEM experiment close to the MMT if the local WF varies within the sample, can be simplified by complementing experiments with ray-tracing simulations.  

The minima and maxima in \dWFapp\ shown in \fig{sim-WF} manifest as apparent WF depression and maxima around patches of different materials in experimental results (e.g., the blue lines surrounding WF islands in Fig.\ 5(a) of Ref.\ \cite{Jobst-LEEP}, Fig.\ 3(b) in Ref.\ \cite{Back2017-WF}, or Fig.\ 3(a) in Ref.\ \cite{Cauduro2017-WF}). 
Our analysis presented here shows that those are artifacts purely generated by the in-plane fields due to the WF discontinuity and should not be misinterpreted as true \dWF. 

The total apparent WF difference in \fig{sim-WF} is $\max(\dWFapp(x)) - \min(\dWFapp(x)) \approx 0.8$\,eV and thus much larger than the true $\dWF = 0.5$\,eV. To quantify how much this method overestimates true WF steps, we perform such simulations for various \dWF\ settings. 
Figure \ref{fig:sim-WF}(c) shows that the extracted total \dWFapp\ scales linearly with the true \dWF. A linear fit demonstrates that the canonical method overestimates the WF difference by a factor of $a \approx 1.6$ over a wide range of work functions.

Not only is this artifact in the apparent WF large in energy, but also in its lateral extension. Figure \ref{fig:sim-WF}(b) shows that the length scale over which \dWFapp\ deviates from the true $\dWF = 0.5$\,eV is $\dx \approx 0.3$\um. The size of the WF artifact grows with increasing \dWF, but is already significant for very small WF differences as illustrated in \fig{sim-WF}(d). 

Our analysis clearly reveals that extracting WF differences from the MMT is intrinsically unsuited for measuring \dWF\ close to points where the WF changes since every WF discontinuity intrinsically causes in-plane fields and thus, those \dWFapp\ artifacts. 
Overall, this canonical method greatly overestimates \dWF\ and is particularly inadequate for small-scale objects that are typically investigated in LEEM.

\begin{figure}[!ht]
	\includegraphics[width=\columnwidth]{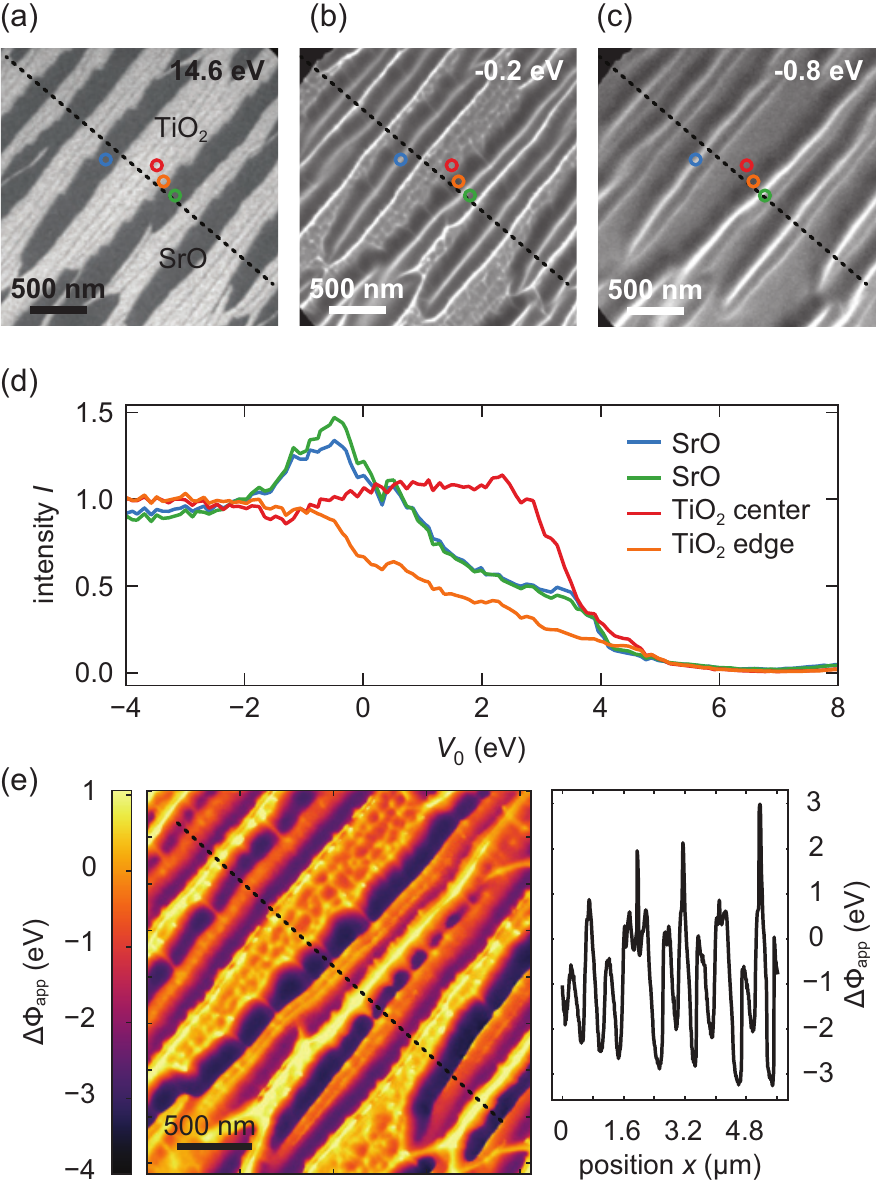}
	\caption{Experimental LEEM data on a mixed-terminated STO surface. 
		(a) A LEEM image at high energy ($\El = 14.6$\,eV) clearly reveals the \tio-terminated areas (bright) and the \sro-terminated parts.
		(b, c) At lower \El, close to mirror mode, dark and bright fringes obscure the exact shape of the domains. 
		(d) The shape of IV-curves close to mirror mode depends more strongly on the position where it is taken than on which termination it is taken from.
		(e) Extracting a work function map by fitting Eq.\ (\ref{eq:error-function}) to every pixel yields very large \dWFapp. Those fits are dominated by the field-induced artifact discussed in this paper. As such, the extracted numbers do not reflect the actual \dWF\, but can easily be mistaken for it.}
	\label{fig:STO}
\end{figure}

\subsection{Combining simulations and LEEM measurements}
In the following, we demonstrate on an experimental data set how extreme this overestimation can be and introduce a more robust way to extract the WF difference between two materials by comparing measured LEEM data to simulations. 

For this purpose, we use mixed-terminated STO as well-defined test sample that exhibits nearly parallel stripes of different WF. The samples are prepared by annealing commercial STO (100) single crystals from Crystec GmbH in air for 12\,h following the recipe described in Ref.\ \cite{Bachelet2009-atomically}. 
Upon this preparation, on approximately half of the area, large \sro-terminated domains form at the otherwise \tio-terminated surface. 
The WF difference between the two terminations is predicted to be large ($\dWF = 1.37$\,eV in Ref.\ \cite{Mrovec2009-Schottky-barrier} and $\dWF = 3.15$\,eV in Ref.\ \cite{Jacobs2016-WF-calculations}), but experimentally much smaller values of $\dWF = 0.22$\,eV \cite{Minohara2010-termination} and even $\dWF < 0.1$\,eV \cite{Aballe2015-instability} have been observed using photoemission spectroscopy and LEEM, respectively. 

A LEEM micrograph in \fig{STO}(a), recorded at $\El = 14.6$\,eV, shows \tio-terminated areas in bright and \sro-terminated areas in dark. At this high landing energy, imaging electrons are hardly affected by the lateral fields introduced by the WF difference between domains of alternating termination \cite{Kautz-LEEP, Jobst-LEEP}. The image is thus very clear and rich in contrast.
At lower energy, close to the MMT, on the other hand, the shape of the different domains is strongly blurred and bright and dark fringes are clearly visible in \fig{STO}(b) (recorded at $\El = -0.2$\,eV). Those fringes correspond to the field-induced caustics found in the simulations shown in \fig{sim-profile}(a,b). Their exact shape and intensity changes with \El\, but they are well visible deep in mirror mode, e.g., in \fig{STO}(c) at $\El = -0.8$\,eV. 

In other words, the shape of the experimental IV-curves is affected by local in-plane fields around WF steps as discussed above. 
Figure \ref{fig:STO}(d) shows IV-curves measured at different points of the STO surface [marked by circles in \fig{STO}(a,b)] as examples of this strong position dependence. In particular, the shape of the IV-curves depends more strongly on position than on the termination from which it was recorded. 
In fact, the experimental IV-curves in \fig{STO}(d) show either a peak before the MMT or a very gradual decrease of intensity, which have been identified as hallmark signatures of field-induced artifacts in the simulation results in \fig{sim-WF}(a). 
Consequently, following the canonical approach to extract the local WF \WFx\ from fitting Eq.\ (\ref{eq:error-function}) to the MMT at every point, yields very high apparent WF values [\fig{STO}(e,f)]. As we have shown above [cf.\ \fig{STO}(d)], however, those numbers arise purely from the deflection of the electron trajectories, greatly overestimate the true WF difference between the two terminations and are thus virtually meaningless. 

\begin{figure}[!ht]
	\includegraphics[width=\columnwidth]{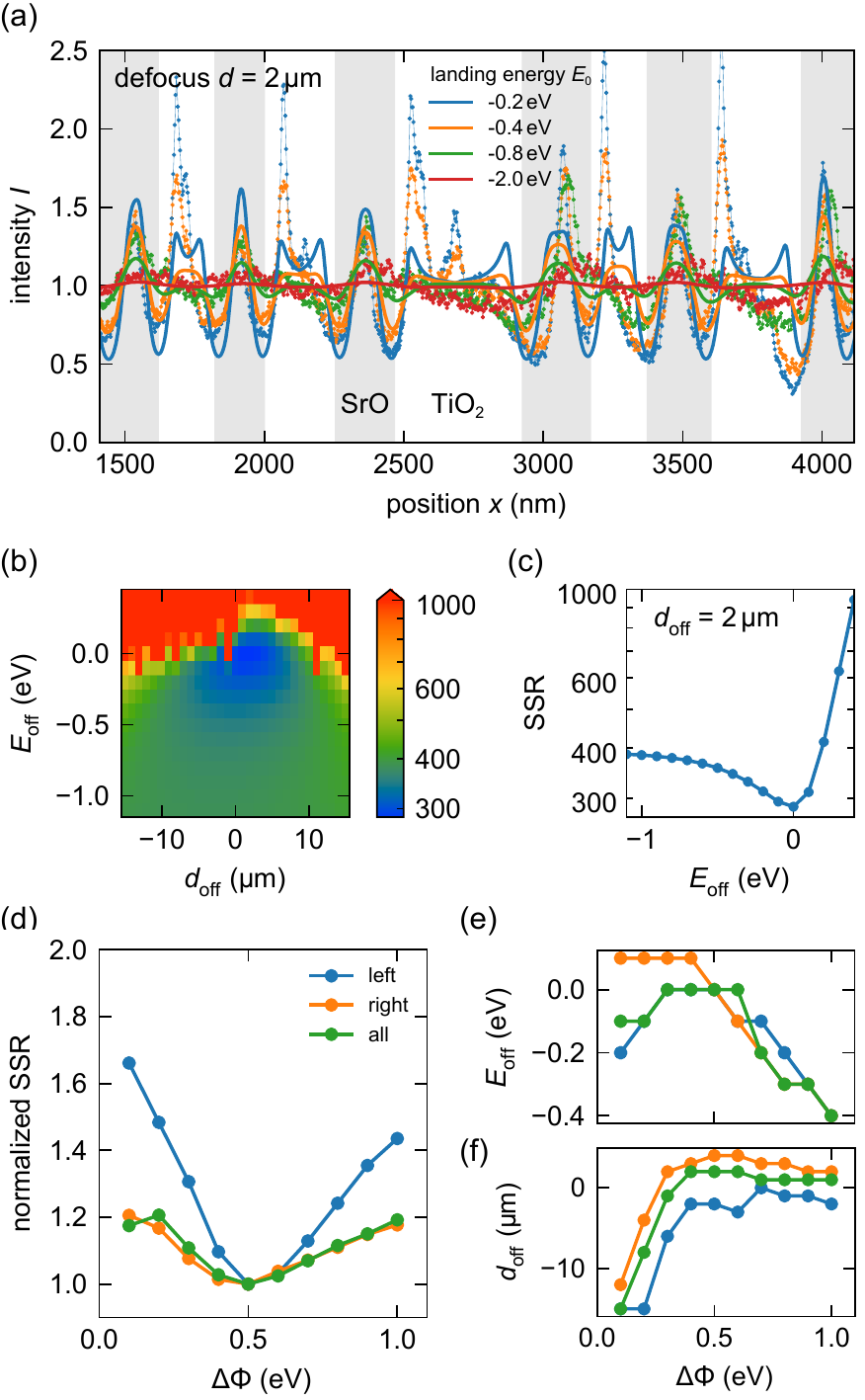}
	\caption{Comparing measured intensity profiles with simulated ones.
		(a) The reflected intensity (dots) measured along the line dashed in \fig{STO}(a,b) shows clear peaks around the \sro-terminated stripes (gray areas). The features are most pronounced close to $\El = 0$ and get smoothed out at higher \El. This behavior is well described by the simulated intensity profile (thicker lines) using $\dWF = 0.5$\,eV.
		(b) The experimental offset in landing energy and defocus can be found by calculating the SSR for various \dOff\ and \Eoff\ using Eq.\ (\ref{eq:SSR}). 
		(c) Cut through the SSR landscape in (b) along the $\dOff = 2$\um\ line. The minimum corresponds to the optimal \dOff, \Eoff\ and SSR.
		(d) The normalized, optimal SSR for different \dWF\ indicates that the data is best described by $\dWF = 0.5$\,eV. Fitting to different parts of the experimental data yields slightly different optimal \Eoff\ (e) and \dOff\ (f).}
	\label{fig:STO-profile}
\end{figure}

A much more robust way to quantify \dWF\ of the two terminations is to compare the experimental data to simulations. To do this, we extract the spacing of \sro- and \tio-terminated stripes from a line profile in \fig{STO}(a) and use them to set up a SIMION simulation with the same geometry. For simplicity, we assume perfectly parallel stripes in the simulation and calculate intensity profiles $I(x)$ perpendicular to those. 
Figure \ref{fig:STO-profile}(a) shows such calculated profiles (thick lines) for $\dWF = 0.5$\,eV for different landing energies below mirror mode ($\El = 0$ is referenced to \tio-termination). The experimental profiles measured along the line in \fig{STO}(a) are shown in \fig{STO-profile}(a) as dots connected with thin lines for the same energies. 
This comparison demonstrates that most of the image contrast can be described already by a $\dWF = 0.5$\,eV which is in stark contrast to the $\dWFapp \approx 5$\,eV found by the canonical method in \fig{STO}(e). In the following, we discuss how we identify the correct value of \dWF\ and how to get an estimate of the confidence interval of this parameter.

As pointed out above, it is difficult to identify focusing condition $d = 0$ and the true landing energy in an experiment close to a WF discontinuity due to the distorted IV-curves. In the simulation, on the other hand, those parameters are well defined. We therefore select a set of profiles for different landing energies \Elexp\ from the experimental data. We cannot yet know the exact value of \Elexp\ but we know that the difference between different \Elexp\ is correct and \Elexp\ and the true \El\ can only differ by a constant offset $\Eoff = \El - \Elexp$. To determine \Eoff\ we chose experimental profiles that span the full range of profiles from almost flat ones (deep in mirror mode) to ones with pronounced spikes (close to the MMT). 
Next we calculate the sum of squared residuals between the experimental profiles \Iexp\ and the simulated profiles \Isim
\begin{equation}\label{eq:ssr}
\ssr = \sum_{i}\left( \Iexp - \Isim \right)^2,
\end{equation}
for all positions $x_i$ and then add them up for all selected \Elexp
\begin{equation}\label{eq:SSR}
	\SSR = \sum_{\Elexp} \ssr, 
\end{equation} 
with the offset \dOff\ from perfect focus in the experiment. This yields an indicator of the quality of the description of the experimental data by a given set of simulations. In addition to the \dWF\ set during the simulation, \SSR\ depends only on \Eoff\ and \dOff\ and allows us to compare different simulations with the data set to decide which parameters best describe our experiment. 
Figure \ref{fig:STO-profile}(b) shows the calculated SSR for a set of \Eoff\ and \dOff\ indicating a clear minimum around $\dOff=0$ (blue). The cut through \fig{STO-profile}(b) along the best-fitting $\dOff = 2$\um\ shown in \fig{STO-profile}(c) illustrates that this routine yields a robust measure to determine the correct focus and energy scale, which is difficult experimentally.

We use this methodology to compute ideal \Eoff\ and \dOff\ together with the resulting SSR for various \dWF\ (this requires to rerun SIMION for every \dWF\ and to extract image profiles \Isim\ for all the runs). The results are summarized in \fig{STO-profile}(d) and confirms that $\dWF = 0.5$\,eV can, indeed, best describe the measured results in \fig{STO} and \fig{STO-profile}(a). 
A closer look at \fig{STO-profile}(a) reveals that the left ($x < 2700$\,nm) and right side $x > 2700$\,nm) are slightly different. In particular, the maxima in the \sro-terminated areas are more pronounced in the right part of the image even though the \sro\ widths are comparable. 
The simulations describing the left and the right side separately (found by fitting the simulations only to half of the data) consequently differ a bit. While the extracted optimal \Eoff\ is similar, we find a clear difference in the optimal \dOff\ shown in \fig{STO-profile}(e) and (f), respectively. This indicates that the alignment during the experiment was not optimal (e.g., a not perfectly collimated electron beam), causing slightly different focus conditions across the field of view. It is noteworthy, that the defocus values of $d = -2$\um\ and $d = 4$\um\ for left and right half are very small; particularly when taking into account that those values are values in the virtual image plane at $2L$ and thus correspond to a true stage displacement of only $\ds = d/3.2$ \cite{Tromp-defocus}.
Moreover, the extracted value of \dWF\ is robust against those small focusing effects as shown in \fig{STO-profile}(d). 
The main discrepancy between data and simulations is the strong asymmetry of the peaks in the \tio\ areas in the former. We attribute this to a small beam tilt that is not considered in the simulations. 

\section{Summary \& Conclusions}
Samples that contain areas of different WF pose a serious problem for the quantitative interpretation of mirror-mode LEEM data. All WF discontinuities inevitable cause lateral fields that deflect slow, incoming electrons. Those deflections cause imaging artifacts such as caustics around the WF step as well as position-dependent deformations of the LEEM IV-curves. 
In particular, around the MMT, where electrons are particularly slow and thus vulnerable to lateral fields, the IV-curves deviate from the ideal, step-like shape. Fitting an error function to extract the local WF, causes characteristic errors that can be spotted in experimental WF maps as ridges and depressions around WF features. 
Note that here we only consider a WF difference on a flat surface. On real samples WF discontinuities almost always are accompanied by a step in sample height. The image deformation due to this geometric effect will complicate matters further \cite{Kennedy-MM-distortions, Nepijko2001-peculiarities}, but can be neglected for many materials where the substrate steps are below 1\,nm in height.
Moreover, on materials where that exhibit a band gap at the vacuum level, the shape of the LEEM-IVs is changed and the apparent MMT shifted \cite{fujikawa-silver, Flege-IV, Jobst-ARRES, Jobst-ARRES-GonBN}, further complicating the interpretation. 
We showed using ray-tracing simulations that for a single WF step, this canonical treatment leads to an overestimation of the WF difference between the two materials of a factor of $\sim$1.6 compared to the true value.
Moreover, the exact height and shape of those features is strongly dependent on the exact focusing conditions. The lateral extension of those WF artifacts ranges between 100\,nm and 300\,nm for $-0.5\,\text{eV} < \dWF < 0.5\,\text{eV}$. 
For small scale structures, this effect can be even more extreme as we demonstrate on a mixed-terminated STO sample. Here the canonical method finds an apparent WF difference of $\sim$5\,eV, while our refined method yields only 0.5\,eV. This value is close to other experimental findings \cite{Minohara2010-termination, Aballe2015-instability} but much lower than the theoretically predicted values of up to $\sim$3\,eV \cite{Mrovec2009-Schottky-barrier, Jacobs2016-WF-calculations}. Such high values that were calculated for unreconstructed surfaces are energetically unfavorable and might be one of the driving forces behind the formation of the $2\times2$ and the $\sqrt{13}\times\sqrt{13}$R33.7\textdegree\ reconstructions of the \sro\ and \tio-terminated surfaces, respectively \cite{Bachelet2009-atomically, Bachelet2009-self-assembly, Kienzle2011-vacant-site}.
Here, we achieve a more error-tolerant result for the work function difference by simulating the images of the experimentally observed geometry for different \dWF\ and calculating the best fit to the data. 
We show that this method is only slightly affected by smaller focusing imperfections and thus presents a robust method to extract WF differences on nanoscopic samples in LEEM.

\section*{Acknowledgments}
We are grateful to Ruud van Egmond, Douwe Scholma and Marcel Hesselberth for technical assistance and to Vera Janssen and Tobias de Jong for feedback on the manuscript. This work was supported by the Netherlands Organization for Scientific Research (NWO) via a VENI grant (680-47-447, J.J.). C.Y.\ is supported by the China Scholarship Council with grant number 201508110214.

\section*{Data Availability}
The data sets presented here can be downloaded at DOI (will be made available upon acceptance of the manuscript) together with the setup files to run the SIMION simulations.

\section*{Author Contributions}
J.J.\ and L.B.\ performed experiments, simulations and data analysis. C.Y.\ and J.A.\ provided the STO samples. All authors contributed to data interpretation and the writing of the manuscript. 

\printbibliography
\end{document}